\documentclass[12pt]{article}
\pdfoutput=1

\DeclareFontFamily{OT1}{pzc}{}
\DeclareFontShape{OT1}{pzc}{m}{it}{<-> s * [1.10] pzcmi7t}{}
\DeclareMathAlphabet{\mathpzc}{OT1}{pzc}{m}{it}

\newcommand{\wt}[1]{\widetilde{#1}}

\usepackage{stmaryrd}

\usepackage{draft} 
\usepackage[weather]{ifsym}

\usepackage{hyperref}
\usepackage{graphicx,color,subfig}
\usepackage{cite}
\usepackage{mciteplus}
\usepackage{skak}
\usepackage{empheq}
\usepackage{tikz}
\usepackage{bbm}

\usetikzlibrary{calc}
\usetikzlibrary{snakes}
\usetikzlibrary{arrows.meta}
\usetikzlibrary{decorations.pathmorphing}
\usetikzlibrary{decorations.markings}
\usetikzlibrary{bending}
\tikzset{snake it/.style={decorate, decoration=snake}}
\usetikzlibrary{shapes.misc}
\tikzset{cross/.style={cross out, draw=black, minimum size=2*(#1-\pgflinewidth), inner sep=0pt, outer sep=0pt},
cross/.default={1pt}}

\usepackage[T1]{fontenc}
\usepackage{esint}
\usepackage{lmodern}

\def\be#1\ee{\begin{align}#1\end{align}}

\definecolor{dark green}{rgb}{0.7,1,0.64}

\usepackage{listings}
\usepackage{xcolor}

\definecolor{codegreen}{rgb}{0,0.6,0}
\definecolor{codegray}{rgb}{0.5,0.5,0.5}
\definecolor{codepurple}{rgb}{0.58,0,0.82}
\definecolor{backcolour}{rgb}{0.95,0.95,0.92}

\lstdefinestyle{myStyle}{
    belowcaptionskip=1\baselineskip,
    breaklines=true,
    frame=none,
    numbers=none, 
    basicstyle=\footnotesize\ttfamily,
    keywordstyle=\bfseries\color{green!40!black},
    commentstyle=\itshape\color{purple!40!black},
    identifierstyle=\color{blue},
    backgroundcolor=\color{gray!10!white},
    tabsize=2,
}

\usepackage{array}

\begin{document}

\unitlength = .8mm

\begin{titlepage}

\begin{center}

\hfill \\
\hfill \\
\vskip 1cm

\title{Diffeomorphism in Closed String Field Theory}

\author{Ben Mazel, Charles Wang, Xi Yin}

\address{Jefferson Physical Laboratory, Harvard University, 
Cambridge, MA 02138 USA
}

\end{center}

\abstract{We explain how the spacetime diffeomorphism in the classical bosonic closed string field theory (SFT) is represented as $L_\infty$ gauge transformations in weakly curved backgrounds. In particular, we demonstrate the explicit map between covariant spacetime fields and the string fields in the flat-vertex frame to the leading few orders in perturbation theory. We further consider a Batalin-Vilkovisky system of extended string fields that manifests diffeomorphism invariance and allows for extending certain perturbative SFT solutions to the large field regime. 
}

\vfill

\end{titlepage}

\eject

\begingroup
\hypersetup{linkcolor=black}

\tableofcontents

\endgroup

\section{Introduction}

Closed string field theory (SFT) is based on an Batalin-Vilkovisky (BV) action functional constructed as a formal power series in the string field \cite{Zwiebach:1992ie}, and provides a complete treatment of string perturbation theory \cite{Sen:2014pia,  Pius:2014gza, Pius:2014iaa, Sen:2019jpm}. It also allows for describing general perturbative deformations of spacetime backgrounds in string theory as solutions to the SFT equation \cite{Sen:1993kb, Sen:2014dqa, Cho:2018nfn, Mazel:2024alu}. On the other hand, the perturbartive definition of closed SFT appears to pose sharp limitations on its utility in describing general string backgrounds, particularly concerning deformations of spacetimes that involve topology change or the emergence of horizons, even in the approximation of weak curvature and field strengths. The difficulty is rooted in the lack of a formulation of SFT that is manifestly local or diffeomorphism invariant in the target spacetime.

In principle, one expects diffeomorphism to be realized in the closed SFT as a gauge transformation \cite{Ghoshal:1991pu}. However, the latter is implemented through an $L_\infty$-algebra, where the diffeomorphism algebra may only be recognized ``up to homotopy''. At a practical level, the map from deformations of the spacetime metric or other covariant tensor fields to the components of the closed string field depends on the string field frame, or equivalently a choice of a consistent set of string vertices, and is not easy to determine a priori. The goal of this paper is to give a systematic treatment of diffeomorphism in closed SFT and to extend the validity of closed string field solutions to describe general spacetime backgrounds at least in the weak field strength approximation.

To begin with, we consider perturbative string field solutions that describe spacetime backgrounds of small curvature and weak fields strengths. In this case, the string field can be split into its massless and massive components, and the equation of motion may be reduced to that of a massless effective SFT \cite{Sen:2016qap, Erbin:2020eyc}. By explicitly identifying the SFT gauge transformations that correspond to spacetime diffeomorphism and 1-form gauge symmetry associated with the $B$-field to the leading orders in perturbation theory, we determine the dictionary between the covariant spacetime fields and components of the massless string field in the flat-vertex frame introduced in \cite{Mazel:2024alu}. As an application and a consistency check, we construct the perturbative string field solution that describes the deformation of an $\mathbb{R}^3$ factor of the spacetime geometry to $S^3$ with uniform $H$-flux, and recover from the string field description the central charge shift of the corresponding worldsheet CFT.

To extend this analysis to all orders in perturbation theory, we propose a framework in which a family of string fields $\Psi_x$, associated with the Riemann normal coordinates centered at a point $x$ on the spacetime manifold, is considered simultaneously with the diffeomorphism gauge transformations that transport between different Riemann normal coordinates. The SFT equation of motion and the gauge transformations naturally combine into the equations of an extended BV system. This formalism allows for extending the perturbative SFT description of spacetime background deformations to the regime of large fields, possibly with topology change, provided that the curvature and field strengths are weak. We will illustrate this approach through the $\mathbb{R}^3\to S^3$ deformation, where the spectrum of global fluctuations can be analyzed in the SFT framework.

The rest of the paper is organized as follows. In section \ref{sec:gensft} we briefly recap the formulation of classical bosonic closed SFT, the flat-vertex string field frame, the $L_\infty$ structure of gauge transformations, and the massless effective SFT. In section \ref{sec:diffeopert}, we identify the gauge transformations that generate spacetime diffomorphism and 1-form gauge symmetry of the $B$-field, and the map between the covariant metric, $B$-field, and dilaton and the massless string fields in the flat-vertex frame. In section \ref{sec:extendedsft}, we introduce the extended BV system that manifests diffeomorphism gauge transformations on the string fields. As an application, we describe the $\mathbb{R}^3\to S^3$ deformation with $H$-flux in SFT by identifying the perturbative string field solution in section \ref{sec:hfluxex} and analyzing the spectrum of fluctuations in section \ref{sec:sthreespec}. We conclude with a summary and future prospectives in section \ref{sec:conclusion}.

\section{Generalities of classical bosonic closed SFT}
\label{sec:gensft}

We will focus on the classical bosonic closed SFT in this paper, although much of our considerations admit straightforward generalizations to closed and quantum superstring field theory.

The string field $\Psi$ is defined as a vector in the restricted state space $\hat{\cal H}$ of the worldsheet CFT,
\ie
\hat{\cal H} = \left\{ \Psi \in {\cal H}_{\rm m} \otimes {\cal H}_{bc} : ~ b_0^- \Psi = L_0^- \Psi = 0\right\},
\fe
where ${\cal H}_{\rm m}$ and ${\cal H}_{bc}$ stand for the matter and the $bc$ ghost CFT state space respectively, and we adopt the standard notation
\ie
b_0^\pm = b_0 \pm \wt b_0,~~~~ L_0^\pm = L_0 \pm \wt L_0,~~~~ c_0^\pm = {c_0 \pm \wt c_0\over 2}.
\fe
We assign ghost number 1 to $c, \wt c$ and ghost number $-1$ to $b,\wt b$. The BV action functional takes the form
\ie
S_{BV}[\Psi] = - {1\over g_s^2} \left[ {1\over 2} \langle \Psi | c_0^- Q_B |\Psi\rangle + \sum_{n=3}^\infty {1\over n!} \{\Psi^{\otimes n} \} \right],
\fe
and the string field equation of motion takes the form
\ie\label{eom}
E[\Psi] \equiv Q_B \Psi + \sum_{n=2}^\infty {1\over n!} [\Psi^{\otimes n}] = 0.
\fe
The string field vertex  $\{\Psi^{\otimes n}\}$ and the string bracket $[\Psi^{\otimes n}]$ are related by
\ie
\langle \Phi | c_0^- | [\Psi^{\otimes n} ] \rangle  = \{\Phi\otimes \Psi^{\otimes n}\},
\fe
for all $\Phi, \Psi\in \hat{\cal H}$. We refer to \cite{Mazel:2024alu} for further details of conventions and the construction of string vertices and brackets.

A choice of string field frame is equivalent to a choice of string vertices or string brackets, subject to the BV master equation. At the level of the equation of motion (\ref{eom}), it is possible to work with string brackets that do not come from totally symmetric string vertices, such as the flat vertex defined in \cite{Mazel:2024alu} which is particularly convenient for analyzing perturbative solutions.\footnote{Note that a nontrivial field redefinition is required to convert string fields in the flat vertex frame to those appearing in an action which necessarily involve totally symmetric string vertices.} In particular, the 2-string field bracket in the flat vertex frame can be written as
\ie
[\Psi^{\otimes 2}] = b_0^- \mathbb{P}^- r_0^{-L_0^+}\left( \Psi(-z_0) \Psi(z_0) \right),
\fe
where $r_0, z_0$ are constants that obey $r_0 \geq |z_0|+1$, $|z_0|\geq 1$, $\mathbb{P}^-$ is defined as the projector onto $L_0^-=0$ states, and the product $\Psi(-z_0) \Psi(z_0)$ is understood in the sense of OPE.

\subsection{$L_\infty$ gauge transformations}

The infinitesimal gauge transformation of the string field $\Psi$ takes the form
\ie\label{gauge}
\delta_\Lambda\Psi = Q_\Psi\Lambda = Q_B \Lambda + \sum_{n=1}^\infty {1\over n!} [\Psi^{\otimes n} \otimes \Lambda].
\fe
While the physical string field $\Psi$ is Grassmann-even with ghost number 2, the string field $\Lambda$ is Grassmann-odd with ghost number 1. Note that the equation of motion $E[\Psi]$ (\ref{eom}) transforms under (\ref{gauge}) as
\ie
\delta_\Lambda E[\Psi] &\equiv E[\Psi + \delta_\Lambda\Psi] - E[\Psi]
= Q_B \delta_\Lambda\Psi + \sum_{n=1}^\infty {1\over n!} [\Psi^{\otimes n}\otimes\delta_\Lambda\Psi]
\\
&= \sum_{n=1}^\infty {1\over n!} \left(Q_B [\Psi^{\otimes n}\otimes\Lambda] + [\Psi^{\otimes n}\otimes Q_B\Lambda] + \sum_{m=1}^\infty {1\over m!} [\Psi^{\otimes n} \otimes [\Psi^{\otimes m}\otimes \Lambda]] \right)
\\
&= - \left[ e^{\otimes \Psi} \otimes E[\Psi]\otimes \Lambda \right] ,
\fe
where we have used the notation $e^{\otimes \Psi}  = \sum_{n=0}^\infty {1\over n!} \Psi^{\otimes n}$. 

The commutator between a pair of infinitesimal gauge transformations, generated by $\Lambda_1$ and $\Lambda_2$, takes the form \cite{Hohm:2017pnh}
\ie{}
& (\delta_{\Lambda_1} \delta_{\Lambda_2} - \delta_{\Lambda_2} \delta_{\Lambda_1}) \Psi
\\
&= [e^{\otimes \Psi} \otimes Q_B\Lambda_2\otimes \Lambda_1]+   [e^{\otimes \Psi} \otimes [e^{\otimes \Psi} \otimes \Lambda_2] \otimes \Lambda_1] - (1\leftrightarrow 2)
\\
&= Q_\Psi \left[ e^{\otimes \Psi} \otimes\Lambda_1\otimes\Lambda_2 \right] +  \left[ e^{\otimes \Psi} \otimes\Lambda_1\otimes\Lambda_2 \otimes E[\Psi]\right].
\fe
When $\Psi$ is a solution to the SFT equation, i.e. $E[\Psi]=0$, we have
\ie
(\delta_{\Lambda_1} \delta_{\Lambda_2} - \delta_{\Lambda_2} \delta_{\Lambda_1})\Psi = \delta_{[\Lambda_1, \Lambda_2]_\Psi}\Psi,
\fe
where
\ie\label{psicomm}
[\Lambda_1, \Lambda_2]_\Psi \equiv  \left[ e^{\otimes \Psi} \otimes\Lambda_1\otimes\Lambda_2 \right].
\fe
In particular, the commutation relations of gauge transformations depend explicitly on $\Psi$.

\subsection{The massless effective SFT}

For analyzing solutions to the SFT equation, it will be convenient to split the string field $\Psi$ into the ``massless'' and ``massive'' components,
\ie
\Psi = W + (1- \mathbb{P}_\varepsilon^+) \Psi,~~~~ W \equiv \mathbb{P}_\varepsilon^+\Psi,
\fe
where $\mathbb{P}_\varepsilon^+$ is the projector from $\hat{\cal H}$ to its subspace of $\hat{\cal H}$ with $|L_0^+|\lesssim {\cal O}(\varepsilon)$. In the weak field strength or slow-varying field approximation, we will take $\varepsilon\sim \A'/L^2$, where $L$ is the length scale of field fluctuation. With this understanding, we will henceforth denote $\mathbb{P}_\varepsilon^+$ simply by $\mathbb{P}^+$.

Working in a relaxed version of the Siegel gauge
\ie\label{relaxedsiegel}
b_0^+ (1-\mathbb{P}^+) \Psi = 0,
\fe
we can solve $(1-\mathbb{P}^+) \Psi$ from (\ref{eom}) by iterating the formula 
\ie\label{psiiter}
(1-\mathbb{P}^+) \Psi = - {b_0^+\over L_0^+}  \sum_{n+m\geq 2} {1\over n! m!} (1-\mathbb{P}^+)[W^{\otimes n} \otimes ((1-\mathbb{P}^+)\Psi)^{\otimes m}].
\fe
The SFT equation can thus be reduced to an equation for the massless string field $W$ of the form \cite{Erbin:2020eyc}
\ie\label{weom}
Q_B W + \sum_{n=2}^\infty {1\over n!} [W^{\otimes n}]' = 0,
\fe
where the massless effective string bracket $[\cdot]'$, obtained by ``integrating out'' the massive string fields as
\ie{}
& [W^{\otimes 2}]' =   \mathbb{P}^+ [W^{\otimes 2}],
\\
& [W^{\otimes 3}]' = \mathbb{P}^+ [W^{\otimes 3}] - 3  \mathbb{P}^+ \left[W\otimes ({b_0^+\over L_0^+} (1-\mathbb{P}^+) [W^{\otimes 2}])\right] ,
\fe
etc., still respect the BV master equation \footnote{The homotopy algebra structure underlying massless effective SFT was studied in \cite{Kajiura:2003ax, Koyama:2020qfb, Arvanitakis:2020rrk}.}
\ie
Q_B [W^{\otimes n}]' = - n [Q_B W\otimes W^{\otimes (n-1)}]' - \sum_{\ell=1}^{n-2} {n\choose \ell} \left[ W^{\otimes \ell}\otimes [W^{\otimes (n-\ell)}]' \right]'.
\fe

A residual gauge transformation $\Lambda$ is such that the relaxed Siegel gauge condition (\ref{relaxedsiegel}) is maintained, namely
\ie\label{residual}
b_0^+ (1-\mathbb{P}^+) \delta_\Lambda\Psi=0.
\fe
Splitting $\Lambda$ into massless and massive components,
\ie
\Lambda = \Omega + (1-\mathbb{P}^+)\Lambda,~~~~ \Omega \equiv \mathbb{P}^+ \Lambda,
\fe
we can further impose the relaxed Siegel condition on $\Lambda$ itself, namely
\ie
b_0^+ (1-\mathbb{P}^+) \Lambda = 0.
\fe
The massive component of $\Lambda$ is then solved from (\ref{residual}) by iterating
\ie
(1-\mathbb{P}^+)\Lambda &= -  \sum_{n=1}^\infty {1\over n!} {b_0^+\over L_0^+}(1-\mathbb{P}^+)[(W+(1-\mathbb{P}^+)\Psi)^{\otimes n}\otimes (\Omega + (1-\mathbb{P}^+)\Lambda)],
\fe
where $(1-\mathbb{P}^+)\Psi$ is determined by $W$ through (\ref{psiiter}).
The resulting gauge variation of the massless string field $W$,
\ie
\delta_\Omega W &\equiv \mathbb{P}^+(\delta_\Lambda\Psi)
\\
&= Q_B \Omega + \sum_{n=1}^\infty {1\over n!} \mathbb{P}^+[ (W+ (1-\mathbb{P}^+) \Psi)^{\otimes n}\otimes (\Omega + (1-\mathbb{P}^+) \Lambda)]
\\
&= Q_B \Omega + \sum_{n=1}^\infty {1\over n!} [W^{\otimes n}\otimes \Omega]',
\fe
takes a form identical to (\ref{gauge}) except with the string fields and the brackets replaced by their massless counterparts.

\section{Diffeomorphism in SFT: leading order analysis}
\label{sec:diffeopert}

We consider a worldsheet matter CFT consisting of $d$ free bosons $X^\mu$ and an auxiliary CFT $\mathbb{M}$ of central charge $26-d$, and will restrict to string fields that involve only the identity Virasoro module of $\mathbb{M}$. We will inspect the gauge transformations of the massless string field $W$ and compare with the expected transformations of the covariant $d$-dimensional massless fields of bosonic string theory under diffeomorphism. This will lead to a systematic identification between components of the string field and the covariant fields (namely the metric, the $B$-field, and the dilaton), order by order in perturbation theory in the slow-varying field approximation.

\subsection{The identification of metric and dilaton}
\label{sec:metricdil}

We begin by considering a worldsheet-parity even massless string field of ghost number 2, whose general form is
\ie\label{wsfield}
W = {1\over 2} c\wt c  h_{\mu\nu}(X) \partial X^\mu \bar\partial X^\nu + \phi(X) {1\over 2} (c\partial^2 c - \wt c \bar\partial^2 \wt c) + A_\mu(X) c_0^+ (c \partial X^\mu - \wt c \bar\partial X^\mu),
\fe
where $h_{\mu\nu}(x)$ is symmetric with respect to $(\mu\nu)$, and $h_{\mu\nu}(x), \phi(x), A_\mu(x)$ are assumed to be slow-varying functions on $\mathbb{R}^d$. Note that we will not impose the Siegel gauge condition on $W$. The relevant gauge transformation on $W$ is generated by a ghost number 1 worldsheet-parity even string field
\ie
\Omega = V_\mu(X) {1\over 2} (c\partial X^\mu - \wt c \bar\partial X^\mu),
\fe
of the form
\ie\label{deltome}
\delta_\Omega W = Q_B \Omega + [W\otimes \Omega]' + \cdots,
\fe
where $\cdots$ stands for higher order terms in $W$.
Using the formulae\footnote{Throughout this paper the free boson field $X^\mu$ is defined in the $\A'=2$ convention.}
\ie{}
& L_0 f(X) = \wt L_0 f(X) = - {1\over 2} \partial_\mu \partial^\mu f(X),
\\
& L_1 f(X) \partial X^\mu = - \partial_\mu f(X),
\fe
the zeroth order gauge variation of $W$ on the RHS of (\ref{deltome}) is evaluated as
\ie\label{firstomw}
Q_B\Omega &= - {1\over 2} c \wt c (\partial_\mu V_\nu(X)+\partial_\nu V_\mu(X)) \partial X^\mu \bar\partial X^\nu
+ {1\over 2}\partial_\mu V^\mu(X) {1\over 2} (c\partial^2c - \wt c \bar\partial^2\wt c) 
\\
&~~~
- {1\over 2} \partial_\nu\partial^\nu V_\mu(X)  c_0^+ (c\partial X^\mu - \wt c \bar\partial X^\mu).
\fe
Working in the flat-vertex frame of \cite{Mazel:2024alu}, the 2-string bracket appearing on the RHS of (\ref{deltome}) is evaluated as
\ie\label{secondomw}
& [W\otimes \Omega]' = \mathbb{P}^+ [W\otimes \Omega] = b_0^-\mathbb{P}^- \mathbb{P}^+ \left( W(-z_0) \Omega(z_0) \right)
+ \cdots
\\
&= {1\over 2} c \wt c \partial X^\mu \bar\partial X^\nu  \Big[ - V^\rho(X)\partial_\rho h_{\mu\nu}(X) + {1\over 4} V^\rho(X) ( \partial_\mu h_{\rho\nu}(X) + \partial_\nu h_{\rho\mu}(X) ) 
\\
&~~~~~~ + {1\over 2} (\partial^\rho V_\mu(X) h_{\rho\nu}(X)+\partial^\rho V_\nu(X) h_{\rho\mu}(X)) - {1\over 4}( \partial_\mu V^\rho(X) h_{\rho\nu}(X)+ \partial_\nu V^\rho(X) h_{\rho\mu}(X))+\cdots \Big]
\\
&~~~ + \Big[ -  {3\over 4} V^\mu(X) \partial_\mu\phi(X) - {1\over 4} V^\mu(X) A_\mu(X) +\cdots \Big] {1\over 2} (c\partial^2 c - \wt c \bar\partial^2 \wt c) 
\\
&~~~ + \Big[ - {3\over 4} V^\nu \partial_\nu A_\mu + {3\over 4} A^\nu \partial_\nu V_\mu +{1\over 4} V^\nu \partial_\mu A_\nu - {1\over 4} A_\nu \partial_\mu V^\nu
 +\cdots \Big] c_0^+ (c\partial X^\mu - \wt c \bar\partial X^\mu).
\fe
Here $\cdots$ stands for terms that involve more $X$-derivatives which are suppressed in the slow-varying field approximation.
Combining (\ref{firstomw}), (\ref{secondomw}), (\ref{deltome}), we can write
\ie\label{delomegaw}
\delta_\Omega W =  {1\over 2} c\wt c  \delta_\Omega h_{\mu\nu}(X) \partial X^\mu \bar\partial X^\nu + \delta_\Omega \phi(X) {1\over 2} (c\partial^2 c - \wt c \bar\partial^2 \wt c) + \delta_\Omega A_\mu(X) c_0^+ (c \partial X^\mu - \wt c \bar\partial X^\mu),
\fe
where the gauge variation of $h_{\mu\nu}$ is in particular given by
\ie\label{omegah}
\delta_\Omega h_{\mu\nu} &= - \partial_\mu V_\nu - \partial_\nu V_\mu - V^\rho \partial_\rho h_{\mu\nu} + {1\over 4} V^\rho(\partial_\mu h_{\rho\nu} + \partial_\nu h_{\rho\mu}) 
\\
&~~~ + {1\over 2} (\partial^\rho V_\mu h_{\rho\nu} + \partial^\rho V_\nu h_{\rho\mu})- {1\over 4} (\partial_\mu V^\rho h_{\rho\nu} + \partial_\nu V^\rho h_{\rho\mu}) +\cdots
\\
&= (\delta_\mu^\rho - {1\over 2} h_\mu{}^\rho)( \delta_\nu^\sigma -{1\over 2} h_\nu{}^\sigma)(-\nabla_\rho^{(h)} \varepsilon_\sigma -\nabla_\sigma^{(h)} \varepsilon_\rho ) +\cdots.
\fe
In the last line, we have expressed the result in terms of the vector field
\ie\label{epsvec}
\varepsilon^\mu = V^\mu - {1\over 4} h^\mu{}_{\rho} V^\rho +\cdots 
\fe
and the covariant derivative $\nabla^{(h)}$ defined with respect to the metric $\delta_{\mu\nu} + h_{\mu\nu}$, namely
\ie
-\nabla_\rho^{(h)} \varepsilon_\sigma -\nabla_\sigma^{(h)} \varepsilon_\rho = -\partial_\mu \varepsilon_\nu -\partial_\nu \varepsilon_\mu - \varepsilon^\rho \partial_\rho h_{\mu\nu} - \partial_\mu \varepsilon^\rho h_{\rho \nu} - \partial_\nu \varepsilon^\rho h_{\rho \mu} +\cdots.
\fe
(\ref{omegah}) can be equivalently expressed as
\ie\label{omegahtwo}
\delta_\Omega (h_{\mu\nu} + {1\over 2} h_{\mu\rho} h^\rho{}_\nu) = -\nabla_\mu^{(h)} \varepsilon_\nu -\nabla_\nu^{(h)} \varepsilon_\mu  +\cdots,
\fe
where we have omitted higher order terms in $h_{\mu\nu}$ as well as terms involving more derivatives that are suppressed in the slow-varying field approximation. The RHS of (\ref{omegahtwo}) now takes the form of the variation of a Riemannian metric tensor under the infinitesimal diffeomorphism generated by the vector field $\varepsilon^\mu$.
This suggests that the covariant spacetime metric $G_{\mu\nu}(x)$ should be identified as 
\ie
G_{\mu\nu} = \delta_{\mu\nu} + h_{\mu\nu} + {1\over 2} h_{\mu\rho} h^\rho{}_\nu +\cdots,
\fe
whereas (\ref{epsvec}) gives the relation between the SFT gauge parameter $V^\mu(x)$ and the diffeomorphism vector field $\varepsilon^\mu(x)$.

Similarly, the gauge variations of $\phi(X)$ and $A_\mu(X)$ appearing on the RHS of (\ref{delomegaw}) are
\ie\label{deltaphia}
& \delta_\Omega \phi = {1\over 2} \partial_\mu V^\mu - {3\over 4} V^\mu\partial_\mu \phi - {1\over 4} V^\mu A_\mu +\cdots,
\\
& \delta_\Omega A_\mu = - {1\over 2} \Box{V}_\mu - {3\over 4} V^\nu \partial_\nu A_\mu + {3\over 4} A^\nu \partial_\nu V_\mu +{1\over 4} V^\nu \partial_\mu A_\nu - {1\over 4} A_\nu \partial_\mu V^\nu +\cdots.
\fe
Writing $h\equiv h_\mu{}^\mu$, it follows from the first line of (\ref{deltaphia}) and (\ref{omegah}) that
\ie\label{deltabigphi}
& \delta_\Omega \big(\phi + {1\over 4} h + {1\over 4} a h^2 + {1\over 4} b h^{\mu\nu} h_{\mu\nu} \big) 
\\
& = - V^\mu \partial_\mu \big(\phi + {1\over 4} h\big) - {1\over 4} V^\mu ( A_\mu - \partial_\mu \phi - {1\over 2}  \partial^\nu h_{\mu\nu} ) +\big ({1\over 8}-b\big) \partial^\mu V^\nu h_{\mu\nu} -  a h \partial_\mu V^\mu +\cdots
\fe
We observe that the second term in the last line vanishes due to the auxiliary field component of the SFT equation of motion
\ie\label{amueom}
A_\mu = \partial_\mu\phi + {1\over 2} \partial^\nu h_{\mu\nu} + \cdots.
\fe
If we further choose $a=0$, $b={1\over 8}$, 
the RHS of (\ref{deltabigphi}) then takes the form of the variation of a scalar field under diffeomorphism, namely 
\ie
\delta_\Omega \Phi(x) = - \varepsilon^\mu(x) \partial_\mu \Phi (x) + \cdots,
\fe
where
\ie\label{dilatonphiid}
\Phi \equiv \phi + {1\over 4} h + {1\over 32} h^{\mu\nu} h_{\mu\nu} +\cdots.
\fe
Note that the difference between $V^\mu$ and $\varepsilon^\mu$, as given in (\ref{epsvec}), only affects higher order terms in this identification. We thus conclude that $\Phi(x)$ (\ref{dilatonphiid}) should be identified with the dilaton field in spacetime.

\subsection{Including the $B$-field}

We now extend the consideration of the previous subsection to include the $B$-field. This amounts to replacing the massless ghost number 2 string field $W$ (\ref{wsfield}) by $W+\wt W$, where $\wt W$ is odd with respect to the worldsheet-parity and takes the form
\ie
\wt W = {1\over 2} c\wt c b_{\mu\nu}(X) \partial X^\mu \bar\partial X^\nu + \wt\phi(X) {1\over 2} (c\partial^2 c + \wt c \bar\partial^2 \wt c) + \wt A_\mu(X) c_0^+ (c \partial X^\mu + \wt c \bar\partial X^\mu).
\fe
Here $b_{\mu\nu}(x)$ is anti-symmetric with respect to $[\mu\nu]$, and $b_{\mu\nu}(x), \wt \phi(x), \wt A_\mu(x)$ are assumed to be slow-varying functions on $\mathbb{R}^d$. 
The relevant gauge transformation is generated by the ghost number 1 string field $\Omega + \wt \Omega$, whose worldsheet-parity odd component $\wt \Omega$ takes the form
\ie\label{wtomgh}
\wt \Omega = K_\mu(X) {1\over 2} (c\partial X^\mu + \wt c \bar\partial X^\mu) .
\fe
We now analyze the gauge variation
\ie\label{omewwg}
\delta_{\Omega + \wt \Omega} (W+ \wt W) &= Q_B \Omega + [W\otimes \Omega]' + [\wt W\otimes \wt \Omega]'
\\
& ~~~ + Q_B \wt\Omega  + [\wt W\otimes \Omega]' + [W\otimes \wt \Omega]' + \cdots \, .
\fe
The three terms in the second line are evaluated as
\ie
Q_B\wt \Omega &=  {1\over 2} c \wt c (\partial_\mu K_\nu(X) - \partial_\nu K_\mu(X)) \partial X^\mu \bar\partial X^\nu
+ {1\over 2}\partial_\mu K^\mu(X) {1\over 2} (c\partial^2c + \wt c \bar\partial^2\wt c) 
\\
&~~~
- {1\over 2} \partial_\nu\partial^\nu K_\mu(X)  c_0^+ (c\partial X^\mu + \wt c \bar\partial X^\mu),
\fe
\ie{}
& [\wt W\otimes \Omega] '=  \mathbb{P}^+ [\wt W\otimes \Omega] 
= {1\over 2} c \wt c \partial X^\mu \bar\partial X^\nu  \Big[ - V^\rho(X)\partial_\rho b_{\mu\nu}(X) + {1\over 4} V^\rho(X) ( \partial_\mu b_{\rho\nu}(X) - \partial_\nu b_{\rho\mu}(X) ) 
\\
&~~~~~~ + {1\over 2} (\partial^\rho V_\mu(X) b_{\rho\nu}(X) - \partial^\rho V_\nu(X) b_{\rho\mu}(X)) - {1\over 4}( \partial_\mu V^\rho(X) b_{\rho\nu}(X) - \partial_\nu V^\rho(X) b_{\rho\mu}(X))+\cdots \Big]
\\
&~~~ + \Big[ -  {3\over 4} V^\mu(X) \partial_\mu\wt\phi(X) - {1\over 4} V^\mu(X) \wt A_\mu(X) +\cdots \Big] {1\over 2} (c\partial^2 c + \wt c \bar\partial^2 \wt c) 
\\
&~~~ + \Big[ - {3\over 4} V^\nu \partial_\nu \wt A_\mu + {3\over 4} \wt A^\nu \partial_\nu V_\mu +{1\over 4} V^\nu \partial_\mu \wt A_\nu - {1\over 4} \wt A_\nu \partial_\mu V^\nu
 +\cdots \Big] c_0^+ (c\partial X^\mu + \wt c \bar\partial X^\mu),
\fe
and
\ie{}
& [ W\otimes \wt \Omega] ' = \mathbb{P}^+ [ W\otimes \wt \Omega] 
={1\over 2} c \wt c \partial X^\mu \bar\partial X^\nu  \Big[  {1\over 4} K^\rho(X) ( \partial_\mu h_{\rho\nu}(X) - \partial_\nu h_{\rho\mu}(X) ) 
\\
&~~~~~~ + {1\over 2} (\partial^\rho K_\mu(X) h_{\rho\nu}(X) - \partial^\rho K_\nu(X) h_{\rho\mu}(X)) - {1\over 4}( \partial_\mu K^\rho(X) h_{\rho\nu}(X) - \partial_\nu K^\rho(X) h_{\rho\mu}(X))+\cdots \Big]
\\
&~~~ + \Big[ -  {1\over 4} K^\mu(X) \partial_\mu \phi(X) - {1\over 4} K^\mu(X) A_\mu(X) +\cdots \Big] {1\over 2} (c\partial^2 c + \wt c \bar\partial^2 \wt c) 
\\
&~~~ + \Big[ - {1\over 4} K^\nu \partial_\nu  A_\mu + {1\over 4}  A^\nu \partial_\nu K_\mu +{1\over 4} K^\nu \partial_\mu  A_\nu - {1\over 4}  A_\nu \partial_\mu K^\nu
 +\cdots \Big] c_0^+ (c\partial X^\mu + \wt c \bar\partial X^\mu).
\fe
The resulting gauge variation of $b_{\mu\nu}$ can be put in the form
\ie\label{deltaoobmn}
\delta_{\Omega+\wt \Omega} b_{\mu\nu} &=  \partial_\mu K_\nu - \partial_\nu K_\mu - V^\rho \partial_\rho b_{\mu\nu} 
\\
&~~~+ {1\over 4} V^\rho(\partial_\mu b_{\rho\nu} - \partial_\nu b_{\rho\mu}) + {1\over 2} (\partial^\rho V_\mu b_{\rho\nu} - \partial^\rho V_\nu b_{\rho\mu}) - {1\over 4} (\partial_\mu V^\rho b_{\rho\nu} - \partial_\nu V^\rho b_{\rho\mu}) 
\\
&~~~+ {1\over 4} K^\rho(\partial_\mu h_{\rho\nu} - \partial_\nu h_{\rho\mu}) + {1\over 2} (\partial^\rho K_\mu h_{\rho\nu} - \partial^\rho K_\nu h_{\rho\mu}) - {1\over 4} (\partial_\mu K^\rho h_{\rho\nu} - \partial_\nu K^\rho h_{\rho\mu}) +\cdots
\\
&= (\delta_\mu^\rho - {1\over 2} h_\mu{}^\rho)( \delta_\nu^\sigma -{1\over 2} h_\nu{}^\sigma)(\partial_\rho \theta_\sigma -\partial_\sigma \theta_\rho - V^\tau\partial_\tau b_{\rho\sigma} - \partial_\mu V^\rho b_{\rho\nu} + \partial_\nu V^\rho b_{\rho\mu} ) 
\\
&~~~ + {1\over 2} b_\mu{}^\rho (\partial_\rho V_\nu + \partial_\nu V_\rho) - {1\over 2} b_\nu{}^\rho (\partial_\rho V_\mu + \partial_\mu V_\rho)  +\cdots
\fe
where $\theta_\mu(x)$ is defined as 
\ie
\theta_\mu = K_\mu + {1\over 4} h_{\mu\rho} K^\rho - {1\over 4} b_{\mu\rho} V^\rho +\cdots
\fe
(\ref{deltaoobmn}) can be equivalently expressed as
\ie\label{dellkb}
\delta_{\Omega+\wt \Omega} \big(b_{\mu\nu} + {1\over 2} h_\mu{}^\rho b_{\rho\nu} + {1\over 2} b_{\mu\rho} h^\rho{}_\nu\big) = \partial_\mu \theta_\nu -\partial_\nu \theta_\mu - V^\tau\partial_\tau b_{\mu\nu} - \partial_\mu V^\rho b_{\rho\nu} + \partial_\nu V^\rho b_{\rho\mu}+\cdots \,  .
\fe
This suggests that we should identify the spacetime $B$-field as
\ie\label{bbida}
B_{\mu\nu} = b_{\mu\nu} + {1\over 2} h_\mu{}^\rho b_{\rho\nu} + {1\over 2} b_{\mu\rho} h^\rho{}_\nu+\cdots,
\fe
so that (\ref{dellkb}) takes the form of the gauge transformation of a 2-form potential, with the 1-form gauge parameter $\theta_\mu(x) dx^\mu$, combined with the diffeomorphism generated by the vector field $\varepsilon^\mu$ (whose difference from $V^\mu$ only affects the higher order terms).

Note that $\wt A_\mu(x)$ is an auxiliary field that is determined by the SFT equation of motion, whereas $\wt\phi(x)$ can be gauged away by adding to $\wt \Omega$ (\ref{wtomgh}) a term of the form $c_0^+ F(X)$.

Turning on the $B$-field also affects the identification of the metric and dilaton field.
Repeating the analysis of the previous subsection but now including the term $[\wt W\otimes \wt \Omega]'$ in the first line of (\ref{omewwg}), we find the following gauge variations of the worldsheet-parity even string field components due to $\wt \Omega$,
\ie
\delta_{\wt \Omega} h_{\mu\nu} &= {1\over 4} K^\rho(\partial_\mu b_{\rho\nu} + \partial_\nu b_{\rho\mu}) 
 + {1\over 2} (\partial^\rho K_\mu b_{\rho\nu} + \partial^\rho K_\nu b_{\rho\mu})- {1\over 4} (\partial_\mu K^\rho b_{\rho\nu} + \partial_\nu K^\rho b_{\rho\mu}) +\cdots
\fe
and
\ie{}
 \delta_{\wt \Omega} \big(\phi + {1\over 4} h + {1\over 32} h^{\mu\nu} h_{\mu\nu} - {3\over 32} b^{\mu\nu} b_{\mu\nu} \big) 
 = -
{1\over 4} K^\mu (\wt A_\mu + \partial_\mu \wt \phi - {1\over 2}\partial^\nu b_{\mu\nu} ) +\cdots,
\fe
where the RHS, up to higher order terms, vanishes by the auxiliary field equation of motion. It follows that the spacetime metric tensor $G_{\mu\nu}(x)$ and the dilaton field $\Phi(x)$ should be identified as 
\ie\label{gphiid}
& G_{\mu\nu} = \delta_{\mu\nu} + h_{\mu\nu} + {1\over 2} h_{\mu\rho} h^\rho{}_\nu + {1\over 2} b_{\mu\rho} b^{\rho}{}_\nu + \cdots,
\\
& \Phi = \phi + {1\over 4} h + {1\over 32} h^{\mu\nu} h_{\mu\nu} - {3 \over 32} b^{\mu\nu} b_{\mu\nu} + \cdots,
\fe
and that the identification of the diffeomorphism vector field (\ref{epsvec}) should be modified to
\ie
\varepsilon^\mu =  V^\mu - {1\over 4} h^\mu{}_{\rho} V^\rho + {1\over 4} b^\mu{}_{\rho} K^\rho  +\cdots .
\fe

\subsection{Example: constant dilaton and uniform $H$-flux}
\label{sec:hfluxex}

It is instructive to inspect the string field solution in the case of vanishing spacetime dilaton, namely setting $\Phi(x)=0$ with the identification (\ref{gphiid}), with a first order $B$-field or $H$-flux, and a second order metric deformation from the flat background. In this case, the worldsheet-parity even and odd components of the massless string field can be written as
\ie\label{wexam}
& W = {1\over 2} c\wt c  h_{\mu\nu}(X) \partial X^\mu \bar\partial X^\nu + \big( - {1\over 4} h(X) + {3\over 32} b_{\mu\nu}(X) b^{\mu\nu}(X)\big) {1\over 2} (c\partial^2 c - \wt c \bar\partial^2 \wt c) 
\\
&~~~~~~  + \big(- {1\over 4} \partial_\mu h(X) + {1\over 2} \partial^\nu h_{\mu\nu}(X)  + {{3}\over 32} \partial_\mu (b^{\rho\sigma} b_{\rho\sigma})(X) \big) c_0^+ (c \partial X^\mu - \wt c \bar\partial X^\mu) + \cdots
\fe
and
\ie\label{wtexam}
 \wt W =  {1\over 2} c\wt c  b_{\mu\nu}(X) \partial X^\mu \bar\partial X^\nu 
 + \cdots.
\fe
We will assume $\Box b_{\mu\nu} = \partial^\nu b_{\mu\nu}=0$, so that $\wt W$ is $Q_B$-closed to first order. The worldsheet-parity even component of the massless SFT equation can be written as
\ie
Q_B W + {1\over 2} [\wt W^{\otimes 2} ] ' + \cdots = 0.
\fe
The leading two terms in this equation evaluate to
\ie
Q_B W &= {1\over 2} c_0^+ c\wt c \big(-\Box h_{\mu\nu} + \partial_\mu \partial^\rho h_{\nu\rho} + \partial_\nu \partial^\rho h_{\mu\rho}
- \partial_\mu\partial_\nu h + {{3}\over 8}  \partial_{\mu}  \partial_\nu (b^{\rho\sigma} b_{\rho\sigma}) \big) \partial X^\mu \bar\partial X^\nu
\\
&~~~ + \big({1\over 2} \Box h -  {1\over 2} \partial^\mu \partial^\nu h_{\mu\nu} - {3\over 32} \Box (b_{\mu\nu} b^{\mu\nu}) -  {{3}\over 32} \Box (b^{\rho\sigma} b_{\rho\sigma}) \big) {1\over 2} c_0^+(c\partial^2 c - \wt c \bar\partial^2 \wt c) + \cdots
\fe
and
\ie{}
& {1\over 2} [\wt W^{\otimes 2}]' = {1\over 2}  b_0^- \mathbb{P}^- \mathbb{P}^+ (\wt W(-z_0) \wt W(z_0)) + \cdots
\\
&= {1\over 4} c_0^+ c\wt c \Big( {1\over 2} \partial_\mu (b^{\rho\sigma}\partial_\nu b_{\rho\sigma})  - \partial_\mu (b^{\rho\sigma}  \partial_\sigma b_{\rho\nu} )
  - \partial_\nu (b^{\rho\sigma} \partial_\rho b_{\mu\sigma}) - H_{\mu\rho\sigma} H_\nu{}^{\rho\sigma} + 2 \partial_\rho b_{\sigma\mu} \partial^\rho b^\sigma{}_\nu
 \Big) \partial X^\mu \bar\partial X^\nu +\cdots
\fe
where we have defined $H_{\mu\nu\rho} \equiv \partial_\mu B_{\nu\rho} + \partial_\nu B_{\rho\mu} + \partial_\rho B_{\mu\nu}$, and $B_{\mu\nu}$ can be identified with $b_{\mu\nu}$ to leading order via (\ref{bbida}). With a first order $b_{\mu\nu}$ and second order $h_{\mu\nu}$, the covariant metric tensor (\ref{gphiid}) can be identified as $G_{\mu\nu} =\delta_{\mu\nu} + h_{\mu\nu} + {1\over 2} b_{\mu\rho} b^\rho{}_\nu + \cdots$. One can then verify that the $c_0^+ c\wt c \partial X^\mu \bar\partial X^\nu$ component of the SFT equation indeed reduces to the covariant equation
\ie\label{eomhbb}
R_{\mu\nu} -  {1\over 4} H_{\mu\rho\sigma} H_\nu{}^{\rho\sigma} + \cdots = 0,
\fe
where $R_{\mu\nu}$ is the Ricci tensor associated with the metric $G_{\mu\nu}$. In particular, a background with uniform $H$-flux is described by the metric and $B$-field of the form
\ie{}
& G_{\mu\nu} = \delta_{\mu\nu} - {3\over 4} \lambda^2 (\delta_{\mu\nu} x^2 - x_\mu x_\nu ) + {\cal O}(\lambda^4),
\\
& B_{\mu\nu} = \lambda \epsilon_{\mu\nu\rho} x^\rho + {\cal O}(\lambda^3).
\fe
The corresponding massless string field is given by
\ie\label{wwtsol}
\mathbb{P}^+\Psi = W + \wt W &= - {\lambda^2 \over 8} c\wt c (\delta_{\mu\nu} X^2 - X_\mu X_\nu) \partial X^\mu \bar\partial X^\nu + {5 \lambda^2\over 32} X^2 (c\partial^2 c - \wt c \bar\partial^2 \wt c) 
\\
&~~~~ + {7\lambda^2 \over 8} X_\mu c_0^+ (c\partial X^\mu - \wt c \bar\partial X^\mu) + {\lambda\over 2} \epsilon_{\mu\nu\rho} X^\rho c\wt c \partial X^\mu \bar\partial X^\nu + \cdots.
\fe
Note that (\ref{wwtsol}) solves the SFT equation at order $\lambda^2$ up to an anti-ghost dilaton term, namely
\ie\label{epsigsa}
E[\Psi] &= Q_B W + {1\over 2} [\wt W^{\otimes 2} ]' + \cdots
\\
&= {\Delta c \over 12} c_0^+ (c\partial^2 c - \wt c \bar\partial^2 \wt c) + {\cal O}(\lambda^4),
\fe
where $\Delta c$ represents a shift of central charge at order $\lambda^2$ \cite{Mazel:2024alu}
\ie
\Delta c &= 12 \cdot  {1\over 2} \big({1\over 2} \Box h -  {1\over 2} \partial^\mu \partial^\nu h_{\mu\nu} - {3\over 32} \Box (b_{\mu\nu} b^{\mu\nu}) -  {{3}\over 32} \Box (b^{\rho\sigma} b_{\rho\sigma})  \big)  =  - 27 \lambda^2.
\fe
This is precisely in agreement with the worldsheet CFT description of the deformed string background as the $SU(2)_k$ WZW model, whose central charge is $c=3-{6\over k} + {\cal O}(k^{-2})$, and the level $k$ is related to the deformation parameter $\lambda$ by\footnote{In the $S^3$ nonlinear sigma model description of the WZW model, the $H$-flux and the curvature radius $R$ are related by $H_{123}= 3\lambda = {8\pi^2 k\over 2\pi^2 R^3} = \sqrt{2\over k}$.} $\lambda = {1\over 3} \sqrt{2\over k}$. As in the setup of \cite{Mazel:2024alu}, to solve the full SFT equation requires turning on additional components of the string field that also effectively deforms the central charge of the auxiliary CFT $\mathbb{M}$, so as to cancel the anti-ghost dilaton term appearing on the RHS of (\ref{epsigsa}) and maintain criticality of the full worldsheet CFT.

\section{Manifesting diffeomorphism in SFT}
\label{sec:extendedsft}

In this section we present a framework for classical closed SFT around a geometric background, in which diffeomorphism invariance is made manifest. The basic idea is to consider simultaneous a family of string fields related by gauge transformations that relate coordinate systems centered at different points on the spacetime manifold, and formulate the SFT equation together with the diffeomorphism gauge transformations as the equation of motion of an extended BV system.

\subsection{Moving between Riemann normal coordinates}

Suppose $\Psi$ is a string field solution that represents a weakly curved spacetime manifold $M_\Psi$. For now we assume a slow-varying dilaton profile $\Phi(x)$, and that the $B$-field vanishes. In a given coordinate system $x^\mu$ on $M_\Psi$, we can view the metric $G_{\mu\nu}(x)$ on $M_\Psi$ as a deformation away from the Euclidean (or Minkowskian) metric, and have determined in section \ref{sec:metricdil} the string field $\Psi$ to leading orders with respect to the metric deformation and the dilaton. We will now consider a family of Riemann normal coordinates $N_x$, centered at each point $x\in M_\Psi$, and a corresponding family of string fields $\Psi_x$ that are related to $\Psi$ by suitable gauge transformations.

To specify the Riemann normal coordinates requires a choice of local frame, as follows. Let $TM_\Psi$ be the tangent bundle of $M_\Psi$, and let $FM_\Psi$ be the corresponding frame bundle. We choose a section of $FM_\Psi$, which amounts to a local orthonormal frame $e_a(x)\equiv e_a{}^\mu(x) \partial_\mu$ that obey $G_{\mu\nu}(x) e_a^\mu(x) e_b^\nu(x) = \delta_{ab}$. The choice of frame also specifies a spin connection $\omega_\mu{}^{ab} = e^a{}_\nu (\partial_\mu e^{b \nu} + \Gamma_{\mu\rho}^\nu e^{b \rho})$, such that the spin-covariant derivative of $e_a(x)$ vanishes.

Let $\varphi_x: T_x M_\Psi \to M_\Psi$ be the geodesic flow map. Namely, given a tangent vector $\eta\equiv \eta^a e_a$, $\varphi_x(\eta)$ is the solution to the geodesic equation beginning at $x$ with initial velocity $\eta$, evaluated at unit time:
\ie{}
& \ddot X^\mu(s) + \Gamma^\mu_{\nu\rho}(X(s)) \dot X^\nu(s) \dot X^\rho(s) = 0,
\\
& X^\mu(0) = x^\mu, ~~~~ \dot X^\mu(0) = \eta^a e_a^\mu, ~~~~ X^\mu(1) = \varphi_x^\mu(\eta).
\fe
We define the Riemann normal coordinates based at $x$ by the coordinate map
\ie
N_x: &~ \mathbb{R}^d \to M_\Psi,
\\
&~ \vec t \mapsto \varphi_x(t^a e_a(x)).
\fe
Given a pair of points $x, y\in M_\Psi$, the Riemann normal coordinates based at $x$ and $y$ are related by the diffeomorphism 
\ie
f_{x,y} = N_y\circ N_x^{-1}.
\fe
In particular, if we take $y$ to be infinitesimally away from $x$, the diffeomorphism $f_{x,y}$ is generated by the vector field
\ie
\wt v_{x, a} = {\partial f^*_{x, N_x(t)} \over \partial t^a}\Big|_{t=0} ,
\fe
where $f^*$ stands for the pullback via $f$, and we have represented the vector field as a linear map on $C^\infty(M_\Psi)$.  We will further define
\ie
 v_{x,\mu} \equiv e^a{}_{\mu}(x) \wt v_{x,a}.
\fe
The obvious integrability property $f_{y,z}\circ f_{x,y} = f_{x,z}$, applied to nearby points $x, y, z$, leads to the following differential equation satisfied by $ v_{x,\mu}$, 
\ie\label{vint}
\partial_{x^\mu}  v_{x,\nu} - \partial_{x^\nu}  v_{x,\mu} + [ v_{x,\mu},  v_{x,\nu}] = 0.
\fe

The infinitesimal diffeomorphism associated to the vector field $ v_{x,\mu}$ should be realized in the SFT as gauge transformation generated by a ghost number 1 string field $\Lambda_{x,\mu}$, such that an $L_\infty$ analog of (\ref{vint}) is obeyed:
\ie\label{lamlambd}
\partial_{x^\mu} \Lambda_{x,\nu} - \partial_{x^\nu} \Lambda_{x,\mu} + [\Lambda_{x,\mu}, \Lambda_{x,\nu}]_\Psi = Q_\Psi \Lambda_{x,\mu\nu}^{(2)},
\fe
where $[\cdot, \cdot]_\Psi$ is defined as in (\ref{psicomm}), and $\Lambda_{x,\mu\nu}^{(2)}$ is a certain ghost number 0 string field.
We can then demand that the family of string field solutions $\Psi_x$ are related by such gauge transformations via
\ie\label{psilam}
{\partial \Psi_x \over \partial x^\mu} + Q_{\Psi_x} \Lambda_{x,\mu} =0.
\fe

\subsection{An extended BV system}

We can unify the conditions (\ref{lamlambd}) and (\ref{psilam}) by enlarging the space of string fields from $\hat{\cal H}$ to $\hat{\cal H}$-valued differential forms on the spacetime manifold $M$. To this end, we introduce a set of $d$ Grassmann-odd variables ${\bf c}^\mu$ with ``new ghost number'' 1, and consider the extended string field
\ie\label{xidef}
{\bf\Xi} = \Psi_x + {\bf c}^\mu \Lambda_{x,\mu} + {1\over 2}  {\bf c}^\mu  {\bf c}^\nu \Lambda^{(2)}_{x,\mu\nu} + \cdots,
\fe
and a new BRST operator
\ie\label{eqn:bold_Q}
{\bf Q} = Q_B - {\bf c}^\mu{\partial\over \partial x^\mu} .
\fe
The extended SFT equation
\ie\label{extsft}
{\bf Q} {\bf \Xi} + \sum_{n=2}^\infty {1\over n!} [{\bf \Xi}^{\otimes n}] = 0
\fe
then reproduces the equation of motion for $\Psi_x$ (\ref{eom}), the diffeomorphism invariance condition (\ref{psilam}, and the integrability condition for diffeomorphism (\ref{lamlambd}) at new ghost numbers 0, 1, and 2 respectively. Note that it suffices to restrict $\bf\Xi$ to total (old + new) ghost number 2, thereby dropping terms involving more than two ${\bf c}$ variables.

Our definition of $\bf\Xi$ and the equation (\ref{extsft}) have thus far made no reference to the Riemannian metric of the spacetime manifold $M$ nor the choice of Riemann normal coordinates. The latter as well as the diffeomorphism vector fields moving between Riemann normal coordinates will be specified as a gauge condition on $\bf\Xi$, as follows. The massless component of the string field $\Psi_x$ takes the form
\ie
\label{eqn:psi_form}
\mathbb{P}^+\Psi_x = {1\over 2} c\wt c  h_{x,ab}(X) \partial X^a \bar\partial X^b + \phi_x(X) {1\over 2} (c\partial^2 c - \wt c \bar\partial^2 \wt c) + A_{x,a}(X) c_0^+ (c \partial X^a - \wt c \bar\partial X^a).
\fe
We will demand that $h_{x,ab}(\xi)$, up to quadratic order in $\xi$, agrees with the metric deformation in a Riemann normal coordinate centered at $x$, 
\ie\label{hnorm}
h_{x,ab}(\xi) = {1\over 3} R_{x,acdb} \xi^c \xi^d + {\cal O}(\xi^3).
\fe
Furthermore, we demand that the massless component of ghost number 1 string field $\Lambda_{x,\mu}$,
\ie
\mathbb{P}^+ \Lambda_{x,\mu} = V_{x,\mu a}(X) {1\over 2} (c\partial X^a - \wt c \bar\partial X^a),
\fe
to generate the shift symmetry by a basis vector at the origin of the Riemann normal coordinate system, namely 
\ie
V_{x,\mu a}(\xi) = e_{a\mu}(x) + {\cal O}(\xi).
\fe
We then expect the equation (\ref{extsft}) to constrain both the $x$ and $\xi$ dependence of $h_{x,ab}(\xi)$, $\phi_x(\xi)$, etc., such that
\ie
G^{(x)}_{ab}(\xi) = \delta_{ab} + h_{x,ab}(\xi) + {\cal O}(\xi^3)~~~~{\rm and}~~~ \Phi(x) = \phi_x(0)
\fe
can be identified as the metric tensor in the Riemann normal coordinate centered at $x$ and the scalar dilaton field respectively.

\subsection{$H$-flux}

We can now consider the system of extended string fields (\ref{xidef}), (\ref{eqn:bold_Q}), (\ref{extsft}) to include nontrivial $B$-field simply by relaxing the assumption on the worldsheet-parity of the string fields $\Psi_x$ and $\Lambda_{x,\mu}$. The massless component of $\Psi_x$ now takes the form
\ie
\label{eqn:psiformb}
\mathbb{P}^+\Psi_x & = {1\over 2} c\wt c  h_{x,ab}(X) \partial X^a \bar\partial X^b + \phi_x(X) {1\over 2} (c\partial^2 c - \wt c \bar\partial^2 \wt c) + A_{x,a}(X) c_0^+ (c \partial X^a - \wt c \bar\partial X^a)
\\
&~~~ +{1\over 2} c\wt c  b_{x,ab}(X) \partial X^a \bar\partial X^b + \wt\phi_x(X) {1\over 2} (c\partial^2 c - \wt c \bar\partial^2 \wt c) + \wt A_{x,a}(X) c_0^+ (c \partial X^a - \wt c \bar\partial X^a) ,
\fe
and we will demand\footnote{This ansatz holds only in the flat-vertex frame.}
\ie\label{bnorm}
& h_{x,ab}(\xi) = \big({1\over 3} R_{x,acdb} + {1\over 18} H_{x,ace} H_{x,bd}{}^{e} \big) \xi^c \xi^d + {\cal O}(\xi^3),
\\
& b_{x,ab}(\xi) = {1\over 3} H_{x,abc}\xi^c + {\cal O}(\xi^2).
\fe
The massless component of $\Lambda_{x,\mu}$ now takes the form
\ie
\mathbb{P}^+ \Lambda_{x,\mu} = V_{x,\mu a}(X) {1\over 2} (c\partial X^a - \wt c \bar\partial X^a) + K_{x,\mu a}(X) {1\over 2} (c\partial X^a + \wt c \bar\partial X^a),
\fe
where the compatibility with (\ref{bnorm}) and the extended equation of motion (\ref{extsft}) requires
\ie
K_{x,\mu a}(\xi) = -{1\over 4} e^b{}_\mu(x) H_{x,b a c} \xi^c + {\cal O}(\xi^2).
\fe

As an example, consider the $\mathbb{R}^3\to S^3$ deformation of section \ref{sec:hfluxex}. We can choose a global orthonormal frame $e_a{}^\mu(x)$ as given by the vector fields that generate the left or right $SU(2)$ isometry of the $S^3$, in which case $\Psi\equiv \Psi_x$, $\wt \Lambda_{a} \equiv e_a{}^\mu(x) \Lambda_{x,\mu}$, and $\wt\Lambda^{(2)}_{ab} \equiv e_a{}^\mu(x) e_b{}^\nu(x) \Lambda^{(2)}_{x,\mu\nu}$ are independent of $x$. 
The equations of motion read
\ie\label{womomeom}
& Q_B W + \sum_{n=2}^\infty {1\over n!} [W^{\otimes n} ]' = 0,
\\
& Q_B \wt \Omega_a +  \sum_{n=1}^\infty {1\over n!} [W^{\otimes n}\otimes \wt\Omega_a]'  =0,
\\
& 2 e_{[a}{}^\mu e_{b]}{}^\nu \partial_\mu e^c{}_\nu\, \wt\Omega_c + [e^{\otimes W}\otimes \wt\Omega_a\otimes \wt\Omega_b]'  = Q_B\wt \Omega_{ab}^{(2)} + \sum_{n=1}^\infty {1\over n!} [W^{\otimes n}\otimes \wt\Omega_{ab}^{(2)}]' ,
\fe
where we have defined $\wt\Omega_a \equiv \mathbb{P}^+ \wt\Lambda_a$ and $\wt \Omega_{ab}^{(2)} \equiv \mathbb{P}^+ \wt \Lambda_{ab}^{(2)}$.
Using
$2e_{[a}{}^\mu e_{b]}{}^\nu \partial_\mu e^c{}_\nu = - 2 e^c{}_\nu e_{[a}{}^\mu \partial_\mu e_{b]}{}^\nu = - \langle e^c, [e_a, e_b]\rangle = -{2\over R} \epsilon_{abc}$, where $R={2\over 3\lambda}$ is the radius of the $S^3$, we see that $\wt \Omega_a$ are the ghost number 1 massless string fields that generate an $SU(2)$ global symmetry. In particular, the symmetry algebra is realized via
\ie\label{omegaalg}
[\wt\Omega_a, \wt\Omega_b]'_W \equiv  [e^{\otimes W}\otimes \wt\Omega_a\otimes \wt\Omega_b]'  
= {2\over R} \epsilon_{abc} \wt\Omega_c + \delta_{\wt\Omega_{ab}^{(2)}} W.
\fe

\subsection{Spectrum of fluctuations}
\label{sec:sthreespec}

In the setup of \cite{Mazel:2024alu}, the spectrum of the deformed matter CFT can be extracted from that of fluctuation string fields of the form
\ie
\Phi_x \in {\cal H}_{\mathbb{A}} \otimes {\cal V}_{1-h,1-\wt h}\otimes {\cal H}_{bc},
\fe
where ${\cal V}_{1-h,1-\wt h}$ is the space of Virasoro descendants of the mock primary $\Upsilon$ of weight $(1-h, 1-\wt h)$. The equations of motion that impose the on-shell condition and covariance are
\ie\label{flucteom}
& Q_{\Psi_x} \Phi_x = 0,
\\
& {\partial \Phi_x\over \partial x^\mu} + \left[ e^{\otimes \Psi_x} \otimes \Lambda_{x,\mu} \otimes \Phi_x\right] = 0.
\fe

For the example of the $\mathbb{R}^3\to S^3$ deformation of section \ref{sec:hfluxex}, we can simplify the analysis of spectrum using the $SO(4)$ isometry of the $S^3$, or its $SU(2)$ subgroup generated by the string field $\wt\Omega_a$ (\ref{omegaalg}). In particular, fluctuation string fields can be organized according to their $SU(2)$ quantum numbers as $\Phi_{x,jm}$, such that $\varphi_{x,jm} \equiv \mathbb{P}^+ \Phi_{x,jm}$ obeys
\ie\label{phijmtransform}
e_a{}^\mu(x) \partial_{x^\mu} \varphi_{x,jm} = - [e^{\otimes W}\otimes \wt\Omega_a\otimes \varphi_{x,jm}]' = -{2\over R} (T_a^{(j)})_{mm'} \varphi_{x,jm'},
\fe
where $T_a^{(j)}$ are the $SU(2)$ generators in the spin $j$ representation, with $j\in {1\over 2} \mathbb{Z}_{>0}$ and $m=-j,-j+1,\cdots, j$. 
To determine which of these representations actually appear in the $Q_{\Psi_x}$-cohomology, we apply (\ref{phijmtransform}) twice, and see that Laplacian in $x$ coordinates agrees with the Casimir to order $R^{-2}$
\be
\Delta_{x} \phi_{x,jm} &= \frac{4}{R^2} (T^{a(j)} T_a^{(j)})_{mm'} \phi_{x,jm'} = -\frac{4 j(j+1)}{R^2} \phi_{x,jm}\\
&= - \frac{(2j)(2j+2)}{R^2} \phi_{x,jm}.
\ee
To match the eigenvalue spectrum of the Laplacian on $S^3$, $\lambda_{\ell m}= -\frac{\ell(\ell+2)}{R^2}$ for  $\ell \in \mathbb{Z}_{>0}$, it must be that all representations of $SU(2)$ carrying $j \in \frac{1}{2} \mathbb{Z}_{>0}$ appear exactly once in the $Q_{\Psi_x}$-cohomology. It is worth emphasizing that this result about the spectrum of fluctuations relies crucially on the fact that $x^\mu$ know about the entire target space, not just the patch seen by expanding in $X^\mu$. This is one significant advantage of the extended BV system.

On the other hand, we can also examine (\ref{flucteom}) order by order in $W$ to determine the weights of the scalar primary operator spectrum in the CFT that describes the $S^3$ background. In particular, the tachyon fluctuations take the form
\ie
\varphi_{x,jm} = c\wt c T_x(X) \Upsilon,
\fe
whereas the massless string fields that generate the $SU(2)$ symmetry are
\ie
\wt \Omega_a = {1\over 2} (c \partial X_a - \wt c \bar\partial X_a) + {\cal O}(R^{-1}).
\fe
To leading order, (\ref{flucteom}) become
\ie\label{flucteomleading}
& 0 = Q_B \varphi_{x,j,m} + \hdots = [c_0 (-\frac{1}{2} \partial_\mu^{(X)} \partial^{(X)\mu} - h) + \wt{c}_0  (-\frac{1}{2} \partial_\mu^{(X)} \partial^{(X)\mu} - \wt{h})]  \varphi_{x,jm}  +\hdots ,
\\
&0 =  {\partial \varphi_{x,jm} \over \partial x^\mu} + \left[ \Omega_{x,\mu} \otimes  \varphi_{x,jm} \right] + \hdots= {\partial \varphi_{x,jm} \over \partial x^\mu} -  \partial^{(X)}_\mu \varphi_{x,jm} +\hdots \, .
\fe
Making use of these equations and the full relationship (\ref{phijmtransform}) it turns out that
\be
h\varphi_{x,jm} = \wt{h}\varphi_{x,jm} =- \frac{1}{2} \partial_\mu^{(X)} \partial^{(X)\mu}\varphi_{x,jm} = - \frac{2}{R^2} (T^{a(j)} T^{(j)}_a)_{mm'} \varphi_{x,jm'},
\ee
i.e.
\be
h =\wt{h} = \frac{j(j+1)}{R^2/2} + O(\frac{1}{R^3})
\ee
matching the expected result $j(j+1)/(k+2)$ at order $R^{-2}$ \footnote{recall that $R = \sqrt{2 k} + O(1)$.}. The next order correction to this formula would require computing 3-string brackets.

\section{Conclusion}
\label{sec:conclusion}

In this paper, we studied the gauge transformations of the string field corresponding to the target space diffeomorphisms, and their implications in the context of classical bosonic string field theory. We summarize the results below, before indicating some future and concurrent directions of research.

It is expected that given a solution $\Psi$ to the SFT equations of motion there is a field identification, dependent on SFT frame, relating the massless components of $\Psi$ to the massless spacetime fields: the physical metric $G_{\mu \nu}$, physical dilaton $\Phi$, and physical $B$-field $B_{\mu \nu}$ of the background described by $\Psi$. Working with the flat vertices of \cite{Mazel:2024alu}, we considered how the massless components of a general string field solution (in the weak field and slowly varying approximation) transform under gauge transformations corresponding to spacetime diffeomorphisms generated by an infinitesimal vector field and to gauge transformations of the $B$-field generated by an infinitesimal 1-form. Analyzing the result gave the key identifications (\ref{bbida}) and (\ref{gphiid}) between the physical fields describing the solution and the massless components of the string field (in the flat vertex frame). It is important to note that at next to leading order the identification between the gauge parameters $\varepsilon^\mu$, $\theta_\mu$ describing the transformations of the physical fields and the components of the ghost number 1  gauge parameter string field $\Omega+\wt{\Omega}$ also involves the components of the string field solution $W+\wt{W}$ itself.

As a basic example, we considered to the second order the perturbative SFT solution that describes the deformation of a flat $\mathbb{R}^3$ into $S^3$ supported by uniform $H$-flux and with vanishing constant physical dilaton. The solution (\ref{wwtsol}) has particular numerical coefficients that are fixed precisely by requiring that the physical fields take the correct form after applying the identification described in the preceding. In principle one could find many solutions to the string field equations of motion that have a similar form, but understanding the relationship between the string field and phyiscal fields is key to identifying the solution which correctly corresponds to the desired physical background. We confirmed that the change in central charge (in the framework of \cite{Mazel:2024alu}) matches the expectation from the $SU(2)_k$ WZW model in the $1/k$ expansion. Similarly to \cite{Mazel:2024alu}, the quantization of $k$ is not visible in the perturbative SFT description.

It is noteworthy that in the deformation of the 1d target space analyzed in \cite{Mazel:2024alu}, a judicious gauge choice and simplifying features of the flat brackets allowed us to compute an all-orders solution for the massless components of the string field in the weak field, slow varying approximation using only the 2-string bracket. This is not the case for the straightforward perturbative analysis of \ref{sec:hfluxex}, and the solution (\ref{wwtsol}) leaves the weak field regime for $X^\mu \sim O(1/\lambda)$; in a sense the solution only sees a patch of the target space around $X^\mu = 0$, and does not detect the topology or full isometries of the resulting target space.

To remedy this problem, we introduced an extended BV system in \ref{sec:extendedsft} that simultaneously considers the family of string field solutions in coordinate systems centered at different points on the spacetime manifold and the gauge transformations that correspond to the diffeomorphisms relating these coordinate systems. By consistently solving this extended system of equations, one can transport the SFT solution along the manifold, and in this way the solution is valid even at large distance away from the original starting point, assuming the curvature remains small. In addition, this extended BV system provides a systematic way to study isometries of the resulting spacetime: 
the string field solution should remain unchanged under a gauge transformation by diffeomorphisms which are isometries of the spacetime the solution represents.
We demonstrate this explicitly for an $SU(2)$ isometry of the $\mathbb{R}^3 \to S^3$ solution. The extended BV system of equations also provides a way to study the fluctuations in the new background, while guaranteeing the fluctuations are appropriately covariant with respect to the isometries of the spacetime. We completed this analysis for the tachyonic modes in the $S^3$ background at leading order.
From the $SU(2)_k$ WZW model, we expect that the allowed values of angular momentum $j \leq \frac{k}{2}$ for the fluctuations should be limited by unitarity. To understand this constraint in the SFT formalism, which is closely tied to the quantization of the $H$-flux, is an interesting open problem.

Various aspects of this paper have been extended to the case of superstring field theory in \cite{ads:fivepaper} to study the SFT deformation of $\mathbb{R}^{10} \to AdS_5\times S^5$ background of type IIB superstrings. It would also be interesting to apply this formalism to a spacetime with horizon, or to understand better the SFT in spacetime with boundary which was recently studied in \cite{Stettinger:2024uus, Firat:2024kxq, Maccaferri:2025orz}.

\section*{Acknowledgements}

We would like thank Minjae Cho, Carlo Maccaferri, Ashoke Sen, and Jakub Vosmera for discussions. This work was initiated during the program ``What is String Theory? Weaving Perspectives Together'' at Kavli Institute for Theoretical Physics (KITP), Santa Barbara. We thank Centro de Ciencias de Benasque Pedro Pascual, Spain, and International Centre for Theoretical Physics, Trieste, Italy, for their hospitality during the course of this work. This work is supported by DOE grant DE-SC0007870, and in part by NSF grant PHY-2309135 to the KITP.

\bibliographystyle{JHEP}
\bibliography{SFT}

\providecommand{\href}[2]{#2}\begingroup\raggedright\begin{thebibliography}{10}

\bibitem{Zwiebach:1992ie}
B.~Zwiebach, {\it {Closed string field theory: Quantum action and the B-V
  master equation}},  {\em Nucl. Phys. B} {\bf 390} (1993) 33--152,
  [\href{http://arxiv.org/abs/hep-th/9206084}{{\tt hep-th/9206084}}].

\bibitem{Sen:2014pia}
A.~Sen, {\it {Off-shell Amplitudes in Superstring Theory}},  {\em Fortsch.
  Phys.} {\bf 63} (2015) 149--188, [\href{http://arxiv.org/abs/1408.0571}{{\tt
  arXiv:1408.0571}}].

\bibitem{Pius:2014gza}
R.~Pius, A.~Rudra, and A.~Sen, {\it {String Perturbation Theory Around
  Dynamically Shifted Vacuum}},  {\em JHEP} {\bf 10} (2014) 070,
  [\href{http://arxiv.org/abs/1404.6254}{{\tt arXiv:1404.6254}}].

\bibitem{Pius:2014iaa}
R.~Pius, A.~Rudra, and A.~Sen, {\it {Mass Renormalization in String Theory:
  General States}},  {\em JHEP} {\bf 07} (2014) 062,
  [\href{http://arxiv.org/abs/1401.7014}{{\tt arXiv:1401.7014}}].

\bibitem{Sen:2019jpm}
A.~Sen, {\it {String Field Theory as World-sheet UV Regulator}},  {\em JHEP}
  {\bf 10} (2019) 119, [\href{http://arxiv.org/abs/1902.00263}{{\tt
  arXiv:1902.00263}}].

\bibitem{Sen:1993kb}
A.~Sen and B.~Zwiebach, {\it {Quantum background independence of closed string
  field theory}},  {\em Nucl. Phys. B} {\bf 423} (1994) 580--630,
  [\href{http://arxiv.org/abs/hep-th/9311009}{{\tt hep-th/9311009}}].

\bibitem{Sen:2014dqa}
A.~Sen, {\it {Gauge Invariant 1PI Effective Action for Superstring Field
  Theory}},  {\em JHEP} {\bf 06} (2015) 022,
  [\href{http://arxiv.org/abs/1411.7478}{{\tt arXiv:1411.7478}}].

\bibitem{Cho:2018nfn}
M.~Cho, S.~Collier, and X.~Yin, {\it {Strings in Ramond-Ramond Backgrounds from
  the Neveu-Schwarz-Ramond Formalism}},  {\em JHEP} {\bf 12} (2020) 123,
  [\href{http://arxiv.org/abs/1811.00032}{{\tt arXiv:1811.00032}}].

\bibitem{Mazel:2024alu}
B.~Mazel, J.~Sandor, C.~Wang, and X.~Yin, {\it {Conformal Perturbation Theory
  and Tachyon-Dilaton Eschatology via String Fields}},
  \href{http://arxiv.org/abs/2403.14544}{{\tt arXiv:2403.14544}}.

\bibitem{Ghoshal:1991pu}
D.~Ghoshal and A.~Sen, {\it {Gauge and general coordinate invariance in
  nonpolynomial closed string theory}},  {\em Nucl. Phys. B} {\bf 380} (1992)
  103--127, [\href{http://arxiv.org/abs/hep-th/9110038}{{\tt hep-th/9110038}}].

\bibitem{Sen:2016qap}
A.~Sen, {\it {Wilsonian Effective Action of Superstring Theory}},  {\em JHEP}
  {\bf 01} (2017) 108, [\href{http://arxiv.org/abs/1609.00459}{{\tt
  arXiv:1609.00459}}].

\bibitem{Erbin:2020eyc}
H.~Erbin, C.~Maccaferri, M.~Schnabl, and J.~Vo\v{s}mera, {\it {Classical
  algebraic structures in string theory effective actions}},  {\em JHEP} {\bf
  11} (2020) 123, [\href{http://arxiv.org/abs/2006.16270}{{\tt
  arXiv:2006.16270}}].

\bibitem{Hohm:2017pnh}
O.~Hohm and B.~Zwiebach, {\it {$L_{\infty}$ Algebras and Field Theory}},  {\em
  Fortsch. Phys.} {\bf 65} (2017), no.~3-4 1700014,
  [\href{http://arxiv.org/abs/1701.08824}{{\tt arXiv:1701.08824}}].

\bibitem{Kajiura:2003ax}
H.~Kajiura, {\it {Noncommutative homotopy algebras associated with open
  strings}},  {\em Rev. Math. Phys.} {\bf 19} (2007) 1--99,
  [\href{http://arxiv.org/abs/math/0306332}{{\tt math/0306332}}].

\bibitem{Koyama:2020qfb}
D.~Koyama, Y.~Okawa, and N.~Suzuki, {\it {Gauge-invariant operators of open
  bosonic string field theory in the low-energy limit}},
  \href{http://arxiv.org/abs/2006.16710}{{\tt arXiv:2006.16710}}.

\bibitem{Arvanitakis:2020rrk}
A.~S. Arvanitakis, O.~Hohm, C.~Hull, and V.~Lekeu, {\it {Homotopy Transfer and
  Effective Field Theory I: Tree-level}},  {\em Fortsch. Phys.} {\bf 70}
  (2022), no.~2-3 2200003, [\href{http://arxiv.org/abs/2007.07942}{{\tt
  arXiv:2007.07942}}].

\bibitem{ads:fivepaper}
M.~Cho, J.~Gomide, J.~Scheinpflug, and X.~Yin, {\it The ads${}_5\times$s${}^5$
  solution of superstring field theory},  {\em to appear}.

\bibitem{Stettinger:2024uus}
G.~Stettinger, {\it {A boundary term for open string field theory}},
  \href{http://arxiv.org/abs/2411.15123}{{\tt arXiv:2411.15123}}.

\bibitem{Firat:2024kxq}
A.~H. F\i{}rat and R.~A. Mamade, {\it {Boundary terms in string field theory}},
   {\em JHEP} {\bf 02} (2025) 058, [\href{http://arxiv.org/abs/2411.16673}{{\tt
  arXiv:2411.16673}}].

\bibitem{Maccaferri:2025orz}
C.~Maccaferri, R.~Poletti, A.~Ruffino, and J.~Vo\v{s}mera, {\it {Boundary Modes
  in String Field Theory}},  \href{http://arxiv.org/abs/2502.19373}{{\tt
  arXiv:2502.19373}}.

\end{thebibliography}\endgroup

\end{document}